# Influences of tongue biomechanics on speech movements during the production of velar stop consonants: A modeling study

**Pascal Perrier[1], Yohan Payan[2], Majid Zandipour[3] and Joseph Perkell[3]**

[1]Institut de la Communication Parlée, UMR CNRS 5009, INPG, Grenoble, France

[2] Laboratoire TIMC, CNRS, Université Joseph Fourier, Grenoble, France

[3]Speech Communication Group, R.L.E., Massachusetts Institute of Technology, Cambridge, Massachusetts, USA

Contact :   Pascal Perrier
ICP, INPG
46 Avenue Félix Viallet
38031 Grenoble Cédex 01
France
e-mail : perrier@icp.inpg.fr
Phone 33 + 476 574 825
Fax : 33 + 476 5704 710

## Abstract

This study explores the following hypothesis: forward looping movements of the tongue that are observed in VCV sequences are due partly to the anatomical arrangement of the tongue muscles, how they are used to produce a velar closure and how tongue interacts with the palate during consonantal closure. The study uses an anatomically based two-dimensional biomechanical tongue model. Tissue elastic properties are accounted for in finite-element modeling, and movement is controlled by constant-rate control parameter shifts. Tongue raising and lowering movements are produced by the model mainly with the combined actions of the genioglossus, styloglossus, and hyoglossus. Simulations of V1CV2 movements were made, where C is a velar consonant and V is [a], [i] or [u]. Both vowels and consonants are specified in terms of targets, but for the consonant the target is virtual, and cannot be reached because it is beyond the surface of the palate. If V1 is the vowel [a] or [u], the resulting trajectory describes a movement that begins to loop forward before consonant closure and continues to slide along the palate during the closure. This pattern is very stable, when moderate changes are made to the specification of the target consonant location. This prediction agrees with data published in the literature. If V1 is the vowel [i], we also observed looping patterns, but their orientation was quite sensitive to small changes in the location of the consonant target. These findings also agree with patterns of variability observed in measurements from human speakers, but it contradicts data published by Houde (1967). Our observations support the idea that the biomechanical properties of the tongue could be the main factor responsible for the forward loops when V1 is a back vowel, regardless of whether V2 is a back vowel or a front vowel. In the [i] context it seems that additional factors have to be taken into consideration in order to explain the observations made on some speakers.

# Introduction

Many studies aimed at understanding the control strategies of speech production have been based on analyses of observable (articulatory or acoustic) speech signals. However, it is well known that comparable observed patterns could be produced by different underlying mechanisms.

For example, Nelson (1983) suggested that speech articulatory movements would be produced with an optimized control strategy aiming at minimizing the jerk (the third derivative of displacement versus time). At the same time, he demonstrated that the velocity profile associated with jerk minimization is bell-shaped and quite similar to the velocity profile of a second order system. Since then, the kinematic properties of speech articulators have been shown to be close to those of a second order dynamical model (see for example Ostry and Munhall, 1985, for tongue movements). The central question is, thus, to know whether these kinematic properties are the result of optimized central control (as implied by Nelson) or whether they are a natural consequence of the biomechanical properties of the speech articulators or whether they are the result of the combination of both effects.

Another example of this nature can be found in the studies initiated by Adams *et al*. (1993). These authors observed that when speaking rate decreases, the number of local maxima observed in the velocity profiles of articulatory movements (so-called velocity peaks) would increase from one or two to several. They suggested that the change from fast to slow movements would imply a drastic modification of the underlying control strategy from a single movement to a sequence of multiple sub-movements. However, a study carried out by McClean and Clay (1995) showed that the variability in the number of velocity peaks observed for an articulatory gesture across speaking rates could be related to the firing rate of motor units, which would naturally vary when velocity changes. Thus, far from being an evidence for a drastic modification of the control, multiple velocity peaks could simply originate in the natural variation of a low-level neurophysiological process. Again, these observations raise the question of the relative influences of, on the one hand, control strategies and, on the other hand, physical, physiological and neurophysiological properties, on the kinematic patterns observed during speech production.

In this context, the present paper proposes an assessment of the potential contribution of the biomechanics to complex articulatory patterns (called "articulatory loops") observed during the production of VCV sequences, where C is a velar stop consonant (Houde, 1967; Mooshammer *et al*., 1995, Löfqvist and Gracco, 2002). The study is based on simulations made with a 2D biomechanical tongue model. After a summary of the main experimental observations of articulatory loops in the literature and their possible explanations, the tongue model will be presented and results of various simulations will be described that contribute to the analysis.

## 1. Background: Experimental evidences of "articulatory loops"

Articulatory looping patterns were described the first time in 1967 by Houde, who analyzed cineradiographic data in a number of V1-[g]-V2 sequences, and noted that "*a distinct forward directed gesture takes place during the closure*" of the consonant. Studying tongue body motions from the trajectories of four radio-opaque markers attached to the mid-line of the tongue of a single speaker, Houde noted: "*When the closure occurs during a forward directed vowel transition (/ugi/, /agi/), … the contact appears to be sustained while sliding along the palate for a distance of up to 6 mm.*" (Houde, 1967, p. 129). In these sequences, the observed sliding movement could easily be interpreted as the consequence of the vowel-to-vowel gesture (oriented from the back to the front). However, such a hypothesis would not be consistent with the other set of observations provided by Houde: "*When the* *palatal closure occurs during a rearward movement of the tongue* […], *in some cases (/i'gagi/) its direction is temporarily reversed. It behaves as if forward movement had been superimposed, during contact, on the main rearward movement of the tongue.*" (Houde, 1967, p. 129). In addition, similar movements patterns were also observed in V1-g-V2 sequences where V1=V2, thus apparently precluding an explanation based only on vowel-to-vowel coarticulation phenomena.

Houde suggested that the forward movement could result from a passive effect of forces generated on the tongue surface by the air pressure behind the contact location. Due to the closure of the vocal tract, the air pressure increases in the back cavity and could push the tongue in the forward direction. "*The direction of the movement during closure is consistent with an increase in oral pressure, and as in the case of labial closures, a compliant element is required in the oral cavity, during the voiced palatal stop in order to sustain voicing. The passive reaction of the tongue may provide that required compliance.*" (Houde, 1967, p. 133)

Since then, many additional observations have been made of such loops (Kent and Moll, 1972; Perkell *et al.*, 1993; Löfqvist and Gracco, 1994; Mooshammer *et al.*, 1995; Löfqvist and Gracco, 2002), and the hypothetical influence of air pressure in the back cavity has been analyzed further. Ohala has suggested that this looping movement could be "*a very marked form of active cavity enlargement and could more than compensate for the other factors which disfavor voicing on velars.*" (Ohala, 1983, p. 200). However the hypothesis of active control of the loops has been seriously questioned by data collected on German speakers by Mooshammer *et al.* (1995). Their subjects produced articulatory loops during the unvoiced stop consonant [k] in [aka] that were even larger than for the voiced consonant [g]. This result clearly does not favor Ohala's hypothesis, but it does not refute the assumption that aerodynamics pressure forces could contribute to the forward movement observed in articulatory loops. Hoole *et al.* (1998) tried to assess the potential effect of the pressure forces quantitatively, by experimentally comparing the production of velar consonants during normal *versus* ingressive speech. Their results revealed forward articulatory loops in both conditions, but their size was significantly reduced in ingressive speech. This result confirms the idea that aerodynamics could influence tongue movements, but, at the same time, it also indicates that other factors, perhaps including biomechanical properties of the tongue, may also contribute to generate the observed loops.

On the other hand, Löfqvist and Gracco (2002), inspired by studies of arm control movement in reaching or pointing tasks, suggested that the curvature of the articulatory trajectories, that is at the origin of the looping patterns, could arise from general motor control principles based on a cost minimization. Such a minimization would mean that the whole trajectory of the tongue would be planned, and that physical factors such as aerodynamics and biomechanics would play no direct role or a minimal role in the trajectory shape.

In this paper, we will explore a totally different hypothesis. Using Payan and Perrier's (1997) tongue model, we will assess the hypothesis that biomechanical factors may be at least partly responsible for the observed looping patterns.

## 2. The Tongue Model

Before giving details about the structure of the model, a short description is provided of the anatomy of the tongue together with a brief overview of the state-of-the-art in the field of the biomechanical modeling of the tongue.

## 2.1. A short description of tongue structure

### 2.1.1. Tongue musculature

A detailed anatomical study of the tongue musculature has been described in Takemoto (2001). Thus, the description given here will only address functional aspects (Perkins and Kent, 1986) that were useful in the design of the 2D biomechanical tongue model. Therefore, it will be limited to muscles for which the main influence can be described in the midsagittal plane, and muscles with fibers oriented mainly in the direction orthogonal to the midsagittal plane will be not presented. Most of the considered muscles are paired, with one on each side of the midsagittal plane; however in the following description, their names are given in singular form. Among the ten muscles that act on the tongue structure, there are three extrinsic muscles that originate on bony structures and insert into the tongue: the *genioglossus*, the *styloglossus* and the *hyoglossus*. They are responsible for the main displacement and shaping of the overall tongue structure (e.g., see Perkell, 1996). Contraction of the posterior fibers of the *genioglossus* produces a forward and upward movement of the tongue body, while its anterior fibers pull the anterior portion of the tongue downward. The *styloglossus* raises and retracts the tongue, causing a bunching of the dorsum in the velar region. The *hyoglossus* retracts and lowers the tongue body. Three additional intrinsic muscles, totally embedded in the tongue structure, contribute to a lesser extent to the sagittal tongue shape. The *superior longitudinal* muscle shortens the tongue, and bends its blade upwards. The *inferior longitudinal* muscle depresses the tip. The *verticalis* fibers depress the tongue and flatten its surface.

### 2.1.2. Tongue innervation

Tongue innervation (carrying its motor supply and its sensory and proprioceptive feedback) doesn't involve the same kind of neural circuitry, as does the control of human limb movements. Whereas human limb muscles are innervated by spinal nerves, the muscles of the vocal tract are innervated by cranial nerves, which have their nuclei in the brain stem. However, most of the principles governing limb motor control also apply to the control of tongue movements. For example, as for the limbs, the efferent commands that are conveyed to tongue muscles (by the hypoglossal nerve) are likely to be modulated by proprioceptive feedback. Indeed, most of the oral mucosa, and particularly the tongue surface, is supplied with several different types of mechanoreceptors, and muscles spindles have been found within the tongue musculature (Cooper (1953); Walker and Rajagopal (1959); Porter (1966); Trulsson and Essick (1997)).

## 2.2. The 2D biomechanical model of the tongue

A number of biomechanical tongue models have been already developed to study speech production (for 2 D models, see Perkell 1974; 1996; Kiritani *et al.*, 1976, Hashimoto and Suga, 1986; Honda 1996, Sanguineti *et al.*, 1997; Payan and Perrier, 1997; Dang and Honda; 1998; for 3D models, see Wilhelms-Tricarico, 1995; Kakita *et al.*, 1985). The tongue model used in the current study represents a significant improvement of Payan and Perrier's 2D tongue model (1997). In this section, the fundamental aspects of the modeling approach are described in detail.

### 2.2.1. Biomechanical structure

An important first choice in modeling tongue structure was to limit the complexity of the model by only representing tongue characteristics that are relevant for speech. For this reason only the muscles mainly active during speech have been incorporated (see below for details). In addition, the tongue description has been limited to the midsagittal plane, in accordance with phonetic classifications of speech sounds that are based either on the position of the highest point of the tongue in this plane (Straka, 1965), or on the position of the vocal tract constriction along the midline going from the glottis to the lips (Fant, 1960; Wood, 1979) This choice is also consistent with the fact that the kinematic data available in the literature describe tongue movements in the midsagittal plane (cineradiographic and electromagnetic recordings).

In order to develop a biomechanical model as close as possible to the morphological and physical characteristics of a given speaker, a native speaker of French, PB, who has already been the subject for a large number of acoustic and articulatory data recordings (by cineradiography, electropalatography, electromagnetography, MRI), was employed as our reference speaker.

The PB vocal tract contours (hard palate, velar region, pharynx and larynx), shape and position of the mandible, the lips and hyoid bone, and the surface contour of the tongue were extracted from a lateral X-ray image of PB's head during a pause in a speech utterance. The corresponding tongue shape is, therefore, considered to be close to the rest position of the articulator.

The tongue is composed of a rich mixture of muscle fibers, glands, connective and fatty tissues, blood vessels and mucosa. However, for a first approximation, only two categories of tissues were modeled: the *passive tissues* and the *active tissues*. The first category includes the mucosa, the connective and fatty tissues, blood vessels and glands; the second category corresponds to muscle fibers. Measurements can be found in the literature showing that the stiffness of muscular tissues increases with muscle activation (Duck, 1990 ; Ohayon *et al.*, 1999). This feature is included in the model, by increasing the stiffness of the elements associated with an active muscle.

The *Finite Element Method* (FEM) was chosen to discretize the partial differential equations that describe continuous tongue deformations. These equations were established according to the laws of the linear mechanics. In addition to a precise description of the continuous, visco-elastic and incompressible properties of a body, the FE method makes it possible, via the notion of *element*, to attribute specific biomechanical properties to individual regions of the structure. This feature is crucial in order to make a distinction between passive and active tissues that constitute the model.

Defining and distributing the elements inside the structure (i.e. the *mesh* definition) was the next stage of the procedure. This is a critical phase that involves a compromise among faithfulness to reality, design complexity, and computation time. Automatic mesh generators might have been used here but were not, mainly because we wanted the finite element model of the tongue to represent its muscular anatomy. For this reason, the element geometries were designed manually, with specific constraints in term of (1) the number of elements and (2) anatomical arrangement of the main muscular components. Ideally, it would be optimal to design an FEM structure where the limits of the tissues and muscles could be mapped exactly into the geometry of the different elements.

Figure 1 presents the results of the manual FE mesh design: 221 nodes (intersections of lines in the figure) define 192 quadratic elements (areas enclosed by lines) located inside the sagittal tongue contour. Tongue attachments to the jaw and hyoid bone were modeled by allowing no displacement of the corresponding nodes, while tongue base support (essentially the effect of the mylohyoid muscle ) was modeled by a reaction force, which prohibits any downward movements of the nodes located between the genioid tubercle of the mandible and the hyoid bone.

-- INSERT FIGURE 1 AROUND HERE --

2.2.2. Muscle model

This section describes the definition of the insertion points and fiber orientations for the selected muscles, and the model of force generation that was adopted.

Asymmetries of tongue shapes in the lateral direction have been found in many experimental studies (see for example Stone et al., 1997). However, to our knowledge, it has not been suggested that assymetries would result from an explicit control. Consequently, our approach models the two symmetrical parts of each tongue muscle pair as a single entity. Only action in the midsagittal plane is considered.

*2.2.2.1. Insertions and directions of the muscle fibers*

– INSERT FIGURE 2 AROUND HERE –

Muscles are represented in the model at two different levels (see Figure 2). First, their action on the tongue body is accounted for by "macro-fibers" that specify the direction of the forces and the nodes of the FE mesh to which the forces are applied. Macro-fibers are muscle-specific aggregations of segments (the bold lines on Figure 2) connecting a number of selected nodes of the FE mesh to one another and to points on the bony structures (hyoid bone, jaw, styloid process). As depicted on Figure 2, each muscle is composed of one to seven macro-fibers, over which the global muscle force is distributed. Muscle force generation is modeled in a functional way according to Feldman's Equilibrium Point Hypothesis of motor control ( Feldman, 1986). More detail will be given below.

Second, since the recruitment of a muscle modifies the elastic properties of the muscle tissues, muscles are also represented in the model by a number of selected elements within the FE structure (gray shaded elements on Figure 2), whose mechanical stiffness increases with muscle activation. Since the model is limited to a 2D geometrical representation of the tongue, the association between elements and muscles, depicted in Figure 2, was made on the basis of a simplified projection of the tongue in mid-sagittal plane. Special attention was devoted to assuring that the definition of the macro-fibers, the geometry of the elements and their assignment to muscles preserve the main properties of tongue muscle anatomy (Netter, 1989; Takemoto, 2001). Note that, because the inferior longitudinalis is a thin muscle, it was represented as a single macro-fiber running from the hyoid bone to the tongue tip.

### 2.2.2.2. *Modeling the generation of muscle force*

To model the generation of muscle force, Feldman's "$\lambda$ model" (Feldman, 1966, 1986), also referred to as the *Equilibrium Point Hypothesis* (EPH), was used. This model, introduced for arm motor control, has been employed by Flanagan *et al.* (1990) and Laboissière *et al.* (1996) in their model of the jaw and hyoid bone complex. The $\lambda$ model reflects the claim that $\alpha$ motoneuron (MN) activation, which generates force, is not centrally controlled, but is the consequence of the interaction between a central command and proprioceptive feedback. Feldman (1986) assumes that the Central Nervous System (CNS) independently acts on the membrane potentials of $\alpha$ and $\gamma$ MNs, in a way that establishes a threshold muscle length, $\lambda$, at which muscle activation starts. As the central command specifies changes in $\lambda$, muscle activation, and hence force, vary in relation to the difference between the actual muscle length and $\lambda$. Moreover, due to reflex damping, this activation also depends on the rate of change of muscle length. Feldman assumes that the nervous system regulates the equilibrium point of the muscle-load system by shifting the central command λ, in the form of changes in the central facilitation of MNs, producing a movement to a new equilibrium position.[1]

In the present model, consistent with the experimental force-length measurements reported by Feldman and Orlovsky (1972) for a cat gastrocnemius muscle, the relation between active muscle force and muscle activation is approximated by an exponential function (see Laboissière *et al*., 1996; Payan and Perrier, 1997; Sanguinetti *et al.,* 1997 for more details).

### 2.2.3. Elastic tissue properties

In the absence of any muscle recruitment, the tongue mesh represents passive tissues. Under these conditions, the model consists in an isotropic linear FE structure, whose biomechanical characteristics were chosen in order to model tissue quasi-incompressibility and to replicate mechanical measurements available in the literature.

Accounting for tissue incompressibility would require measuring tissue deformations in 3D space. This can obviously not be done properly in relation to a planar model. In this context, tongue deformations in the direction orthogonal to the midsagittal plane were assumed to be negligible in comparison to the geometrical changes in this plane ( the so-called *plane strain hypothesis)*. In this case, tissue quasi-incompressibility is equivalent to area conservation and can be modeled with a Poisson ratio value close to 0.5 (Zienkiewicz and Taylor, 1989). This hypothesis is well supported by 3D measurements of tongue deformation during speech production, such as the ultrasound data published by Stone et al., (1997) or the MRI data analyzed by Badin et al. (2002).

The small-deformation framework of the FE method provides an account for stiffness modeling through the definition of the Young's modulus $E$ value, which is assumed to fit the tissue stress-strain relationship (Zienkiewicz and Taylor, 1989). To our knowledge, no data are available in the literature about $E$ value for passive tongue tissues, but measurements are reported for other part of human body. Young's modulus values are estimated around 15 kPa for skin (Fung, 1993), 10 kPa for blood vessels and between 10 and 30 kPa for vocal folds (Min *et al.*, 1994). To a first approximation, and after a number of trials, the value of Young's modulus of passive tongue tissues stiffness was set at $E_{passive}$ = 12.25 kPa. With this value the temporal characteristics of tongue movements are realistic, as compared with data collected on real speakers, and the levels of force generated by the main muscles (GGp, GGa, STY and HYO) are between 0.5 and 1.5 Newton, which seems to be reasonable (Bunton and Weismer, 1994).

As mentioned earlier, when a muscle is activated, its fiber stiffness increases. Measured values for human skeletal muscles have been reported to be 6.2 kPa for muscles at rest, and 110 kPa for the same muscles in a contracted position (Duck, 1990). The stiffness of cardiac muscle has been measured at close to 30kPa at rest, and as high as 300kPa when the muscle is activated (Ohayon *et al.*, 1999). In the framework of the FE method, modeling the global increase of muscle stiffness with activation was made possible by increasing the value of Young's modulus of muscular elements. Thus, the value of Young's modulus varied with muscle activation (between $E_{passive}$ at rest, and $E_{max}$ when the muscle is maximally contracted), while other tongue elements have a constant value of $E$ equal to $E_{passive}$. For the present version of the model, and again after a number of trials, $E_{max}$ was fixed at 100 kPa. Because of the various sizes of the muscles, this maximal value is reached for muscle dependent levels of force. Thus, for example, it is reached for a 2.8 N force for the posterior genioglossus and for a 0.8 N force for the hyoglossus. For a force level corresponding to normal speech conditions (i.e. between 0.5 N and 1.5 N) the Young modulus varies between 40 kPa and 75 kPa.

Finally, these elastic parameters were validated by comparing the deformations of the FE structure induced by each tongue muscle with the deformations observed during real speech. The force developed by each muscle was tuned so that the global level of force produced during a tongue displacement was at a level similar to those measured on human tongue during reiterant speech production (Bunton and Weismer, 1994), and the direction and amplitude of the simulated deformations were verified to be compatible with data measured on PB during speech sequences (Badin *et al.*, 1995).

Figure 3 plots the tongue deformations induced by each modeled muscle. The tongue shapes shown in the figure are similar to those seen in a number of cineradiographic studies of speech movements (e.g., Perkell, 1969, Bothorel *et al.*, 1986). Note, however, that the upward curvature of the tongue generated by the action of the Superior Longitudinalis (lowest panel) is not sufficient when compared to real tongue tip deformations. Alternative implementations for this muscle, such as the one proposed by Takemoto (2001), are currently being tested.

**-- INSERT FIGURE 3 AROUND HERE --**

2.2.4. Implementation of tongue-palate contacts in the biomechanical model

During the production of stop consonants, contacts between the tongue and palate dramatically influence tongue trajectories. Therefore, modeling collisions between the upper tongue contour and the palatal contour is necessary. In the present work, this includes two steps, which aim at: (1) detecting the existence of tongue/palate contact and (2) generating resulting contact forces.

From a theoretical point of view, solving the problem of the contact detection between a solid curved surface and deformable structure is quite complex. However, it is simplified considerably here for two reasons. First, the representation is two-dimensional instead of three-dimensional. Second, the contours delimiting the two bodies in contact (tongue against palate) are represented by points connected by straight lines. Under these simplified conditions, the contact detection problem is reduced to the detection of the intersection between two straight lines.

The force applied to the tongue when contact with the palate occurs was calculated according to a method originally proposed by Marhefka and Orin (1996). It is a so-called *penalty method*, based on a non-linear relationship between the contact force and the positions and velocities of the nodes located on the upper tongue contour in the contact area. The basic principle of this method is as follows. If a node located on the upper tongue contour moves beyond the limits represented by the palate contours, a repulsion force $F$ is generated in order to push this node back, up to the point where inter-penetration is no longer detected. This force, applied to a node of the tongue model that is in contact with the palate, is computed according to the equation (1)

$$\begin{aligned} \vec{F} &= (-\alpha \cdot x^n - \mu \cdot \dot{x} \cdot x^n) \cdot \vec{k} \quad \text{if} \quad x < 0 \\ \vec{F} &= 0 \quad \text{if} \quad x \geq 0 \end{aligned} \quad (1)$$

where $x$ is the inter-penetration distance (always a negative value when contact exists) between the node on the dorsal tongue contour and its orthogonal projection onto the palate contour; $\dot{x}$ is the first time-derivative of the inter-penetration distance; $\alpha$ is a coefficient representing the "stiffness" of the collision (a large $\alpha$ corresponds to hard contact); $\mu$ represents the "damping factor" of the collision; $n$ accounts for the non-linearity; and $\vec{k}$ is a unit vector orthogonal to the palate contour.

As emphasized in equation (1), the *penalty method* first tolerates a slight penetration of the tongue into the palate; then, it generates a force that pushes the node outward until the inter-penetration distance is positive, at which point the contact force vanishes and the node is free to move back toward the palate. The cyclical behavior inherent to modeling contact in this way has a tendency to result in instabilities and oscillations. The parameters $\alpha$, $\mu$ and n have been arbitrarily fixed at *ad-hoc* values ($\alpha = 60$; $\mu = 0.5$; $n = 0.8$), in such a way that, in the VCV simulations, the inter-penetration distance and the amplitude of the potential oscillations remain smaller than a tenth of millimeter.

During contact, the tongue is free to slide along the palate. To our knowledge the viscosity constraining this sliding movement has never been measured. However, since the palate and the tongue are covered with saliva, and since saliva is a fluid that has lubricating properties, it is reasonable to assume that this viscosity factor is negligible as compared to the other damping factors that constrain tongue movement. Consequently, in the current model, no viscosity coefficient is used in the direction that is parallel to the palatal contour.

## 3. Simulations with the tongue model.

This section reports the results of a number of simulations that explore the potential role of biomechanical factors in the production of the looping articulatory patterns. The control of the tongue model is based on the concept that there is a separate target for each of the individual sounds of the sequence. Hence, specific target tongue shapes were first designed both for vowels and consonants, on the basis of data published in the literature for these sounds in similar contexts. Then, an initial set of simulations was generated for [aka], [uku], [iki] and [ika] sequences. In a second set of simulations, the effect on the articulatory trajectories of changes in the consonant target was studied. Finally, the effect of tongue palate interaction was analyzed specifically.

### 3.1. Underlying control of the tongue model during VCV sequences

#### 3.1.1. Target-oriented control for vowels and consonants

As explained in section 2, muscle activations result from interactions between descending central control, specified by the variables $\lambda_i$ (index i referring to the different muscles), and the actual muscle lengths. A set of commands, $\lambda_i$, specifies the position of the tongue at which a stable mechanical equilibrium, also called posture, is reached. Feldman's suggestion is that arm movements are produced from posture to posture. In line with this hypothesis, in the current model, a sequence of discrete control variable values ($\lambda_i$), specifying successive postures, underlies a continuous trajectory through a sequence of phoneme targets. Movements are produced with constant rate shifts of the control variables from the settings of one target to those of the next target.

The phoneme targets represent the ideal goals toward which the tongue moves successively during the articulation of the sequence. For a given phoneme, these goals can vary with the phonetic context, since we also assume that their specification is the result of a higher level planning process that takes into account the sequence as a whole and integrates some optimization principles. The description of this planning process is not part of the present paper (see however Perrier *et al.*, 1996a, and Perkell *et al.*, 2000 for related discussions).

It is important to note that it is assumed that the underlying articulatory control is similar for vowels and consonants. However, the relation between the target specification and the tongue position actually reached differs significantly between these two classes of speech sounds. The specified vowel targets are located ventral to the palate contours; consequently, the corresponding tongue shape can actually be produced if the dynamical and time parameterization of the movement is adjusted appropriately. On the other hand, specified consonant targets are located beyond the palate and can therefore never be reached: they are "virtual" targets. Consequently, the tongue position reached during the production of the consonant is different from the specified one. It is the result of the combined influences of the target command and of contact between tongue and palate (c.f. Perkell *et al.*, 2000). The "virtual" target hypothesis has been suggested by Löfqvist and Gracco for labial (1997) and for lingual (2002) stops, and it is supported by a kinematic comparison of articulatory data collected on German speakers and simulations made with the Payan and Perrier tongue model (Fuchs *et al.,* 2001).

### 3.1.2. Sequencing of the commands

Since the current study is focused on the influence of tongue biomechanics on articulatory paths, temporal as well as vowel-to-vowel coarticulatory effects (Öhman, 1966; Fowler, 1980; Perkell and Matthies, 1992; Matthies *et al.*, 2001) were purposely eliminated by making the following simplifications:

- No account is given, at the level of the articulation, of the differences between voiced and unvoiced consonants (see Löfqvist and Gracco, 1994, for examples of such differences); consequently, a unique articulatory target was used to specify the velar consonant in each vowel context. We arbitrarily refer to this consonant with the phonetic symbol [k].
- Symmetrical temporal patterns have been chosen for the movements toward and away from the consonant.

- The times of onsets and offsets of the motor command shift are the same for all muscles.

### 3.1.3. *Selection of targets for [a], [i], [u] and [k]*

As discussed above, it is assumed that the target command for a phoneme results from a higher-level planning process that takes a set of successive targets into consideration. If this principle is applied strictly, the same phoneme, vowel or consonant, pronounced in two different phonetic sequences should be associated with two different target commands. For example, the target commands for [a] and [k] are likely to be different in the [aka] as compared to [aki] or [uka]. However, it is known that, in V1CV2 sequences, velar consonants are much more influenced by the surrounding vowels than the vowels are influenced by the consonant (Keating, 1993). Consequently, in order to minimize the number of simulations, in this work only the consonant target was assumed to vary as a function of the context. Thus, two different target commands were used for [k], a front one associated with front vowel contexts, and a back one associated with back vowel contexts, while a unique target was associated to each vowel.

These target commands were determined after a number of trials according to the following procedure. Knowing the main influence of each muscle on the tongue shape, we first approximated the muscle commands associated with each sound, by modifying them step by step, starting from the rest position, up to the point where a constriction was formed in the appropriate region in the vocal tract. Then, muscle commands were adjusted around this initial configuration so that tongue contours were reasonably close to data published in the literature for the same sound in similar vowel contexts (Houde, 1967, for English, and, for French, Bothorel *et al.* 1986). To determine the two different muscle command sets for the velar consonant, special attention was given to the location and to the size of the contact region along the palate. Accordingly, for each new trial, simulations of [iki] and [aka] sequences were generated, and the shape of the tongue at different times during the consonantal closure was observed and compare to x-ray data. The evaluation criterion was to qualitatively replicate the differences in tongue shape observed experimentally for similar sequences. The resulting tongue shapes corresponding to the three vowels targets ([i], [a], [u]) are shown in Figure 4, and the virtual targets associated with the two different muscle commands sets for the velar consonant, both in front and back context, are shown in Figure 5.

The target defined for the vowel [i] involves activation of the Posterior Genioglossus (GGp) and, to a much lesser extent, the Styloglossus (STY). For [a], the target was obtained with recruitment of the Hyoglossus (HG) and of the Anterior Genioglossus (GGa). The production of [u] is achieved with recruitment of the STY, and, to a much lesser extent, of the GGp.

For the velar consonant targets, three muscles are activated, namely the STY, the GGp and, to a lesser extent, the Inferior Longitudinalis (IL). The balance between the forces produced by GGp and STY determines the difference between the anterior target and the posterior one. Figure 5 shows the corresponding overall tongue shapes (top and middle panels) and more closely in the palatal region (bottom panel). The highest point of the tongue is higher and more fronted for the anterior virtual target configuration. Under actual conditions, when the tongue is in contact with the palate, this difference induces for the anterior target a lengthening of the contact region towards the front, which is consistent with the observations provided for French stops by Bothorel *et al.* (1986) (pp. 180-181). In all cases the force due to gravity was not taken into account.

**– INSERT FIGURE 4 AROUND HERE –**

INSERT FIGURE 5 AROUND HERE –

### 3.2. Simulation of V1-[k]-V2 sequences

#### 3.2.1. Simulations for symmetrical V-[k]-V sequences

Simulations were first generated for [a], [u], and [i] in a symmetrical vowel context. The timing of the commands, the same for all the sequences, is given in Table I. For [a] and [u] the consonantal target was the posterior one, for [i] it was the anterior one. The trajectories of four nodes located in the palatal and velar regions on the upper contour of the tongue were analyzed for the three sequences (see Figures 6, 7 and 8). For [aka] and [uku] we observe forward looping patterns for the four nodes, with different amplitudes depending on the location of the nodes on the tongue and on the vowels: the loops observed in [a] context are clearly larger than in [u] context. For [iki] the movement is backward during the entire consonantal closure gesture; the size of the horizontal displacement is smaller than in the other two vowel contexts.

-- INSERT FIGURE 6 AROUND HERE –

-- INSERT FIGURE 7 AROUND HERE –

INSERT FIGURE 8 AROUND HERE –

### 3.2.2. Simulations for asymmetrical V1-[k]-V2 sequences

Simulations were also made for asymmetrical sequences, where the vowels preceding and following the consonant were different. The timing of the commands was the same as in the symmetrical VCV simulations.

Special attention was devoted to the [ika] sequence, since Houde (1967) observed in some cases in [i'ga] a reversal of the main rearward movement during the consonant. As for [iki], the anterior target was used for the stop consonant.

Figure 9 shows the trajectories of the same four nodes. It can be observed that in this simulation, no reversal is produced and that the tongue slides continuously backward along the palate for about 2mm during the [k] closure$^2$. This result will be discussed later in Section 3.3.2 in relation to the consonant target location.

**–INSERT FIGURE 9 AROUND HERE –**

Concerning asymmetrical sequences V1-[k]-V2 in general, experimental studies (Houde, 1967; Mooshammer *et al*., 1995, Löfqvist and Gracco, 2002) have systematically shown that sequences with V1=[i] show a much smaller amount of movement during the consonantal closure in comparison with V1=[u] or [a]. Figure 10 shows the trajectory described by a node on the dorsal tongue contour for all the contexts. This node is the second from the back on Figures 6 to 9. In order to see the influence of V1 on the amplitude of the sliding movement during the closure, the results obtained for the same V1 are grouped on the same panel. It can be observed that the size of the loop is determined by the first vowel V1, and that the general trend observed on the experimental data is replicated: if V1=[i] the amplitude of the movement is clearly smaller than in the other cases. However, the differences are not as strong as the measurement provided by Mooshammer *et al.* (1995). These results will also be discussed below in relation to the consonant target location (see section 3.3.3.1).

–INSERT FIGURE 10 AROUND HERE –

### 3.3. Influence of target locations and tongue biomechanics

#### 3.3.1. Analysis of the articulatory trajectories generated in the simulations

Three aspects of the articulatory trajectories warrant more in-depth analysis: (1) the direction (forwards or backwards), (2) the loop curvature and orientation (clockwise or counterclockwise) (3) the amplitude of the movement during the consonantal closure (the size of the loop). These properties will be analyzed separately, in relation to specific aspects of the model used to generate the sequences.

##### 3.3.1.1. *Direction of the paths*

In summary, for [ak] and for [uk], the nodes located in the palatal region describe forward-oriented trajectories, while the movement is backward for sequences [ik] whether the following vowel [a] or [i]. In the case of the vowels [u] and [a], the virtual target position of [k] is located anterior to the vowel targets (see Figures 4 and 5 low panels). For vowel [i], the consonantal target is located slightly posterior to the [i] target (see Figures 4 and 5 top panels). Therefore, it can be concluded that, in the model, the direction of the movement during the V-[k] sequences is defined by the locations of the vowel and the consonant targets relative to each other. This influence of the target locations could also be easily predicted with a simple kinematic model that would be controlled in a target-based manner.

Consequently, the biomechanical properties of the tongue model do not play any role in the determination of the main direction of the movement, i.e. whether it is forward or backward oriented. However, a kinematic model by itself would describe straight paths, and could not account for the fact that "the horizontal and the vertical components of movement towards the target are pursued independently" (Mooshammer *et al.*, 1995, p. 20). Both experimental data and our simulations show this phenomenon, since the trajectories are curved. In our simulations the trajectory shapes are determined by the biomechanical properties of the model as explained below.

### *3.3.1.2. Trajectory curvature and loop orientation*

The control underlying all the simulations presented above is extremely simple: the transition between two successive sounds is based on a linear interpolation between the two associated sets of muscle threshold lengths at the targets. Consequently, the curvature of the articulatory trajectories cannot be a direct consequence of the control itself. This phenomenon is due to the biomechanical properties of the tongue model, i.e. the passive tongue elasticity, the muscle arrangements within the tongue, and the force generation mechanism.

The passive elasticity is taken into account with the Finite Element Method. Thus, the continuous mechanics of tongue tissue is modeled: force acting on a specific part of the tongue has consequences on the whole tongue body. The relations among the strains generated in different parts of the tongue are non-linear and depend on the Finite Element Parameters (Young Modulus and Poisson ratio).

Muscle fiber orientations are not constant during a movement, since some of the muscle insertions are fixed (for example the bony insertion of the styloglossus) while others are moving with tongue tissues (for example the other ends of the styloglossus). As a consequence the directions of all muscle forces change during the movement.

Additional non-linearities are introduced in the force generation due to the use of Feldman's control model. Because the model incorporates the concept of a threshold length a muscle can suddenly become active if its length exceeds the threshold. Moreover, once a muscle is active, the force generated is an exponential function of its length. External forces are generated temporarily during the contact between tongue and palate, which adds another non-linearity.

The combination of all these non-linearities is responsible for the curved aspect of the trajectories. Thus, contrary to Löfqvist and Gracco's (2002) suggestion, it is not necessary to invoke a general optimization principle that would plan the entire trajectory in to explain the trajectory shape.

The variation of the magnitudes and orientations of muscle forces during the movement, as determined by the combination of target commands, which specify the time variation of the threshold muscle lengths, and tongue deformation, which modifies the length and the orientation of the muscle fibers, also contributes to the shape and orientation of the loop. For example, because of the combined actions of the GGp and the STY, for [aka] and [uku], the middle part of the tongue first moves upward and then forward before hitting the palate. It can also be observed that after the consonantal closure for [aka], [uku] and [iki], the first part of the movement toward the vowel is forward oriented, although only slightly so for [iki]. This forward movement is observed even if the subsequent vowel is posterior to the consonant release location, even though the motor commands do not specify movement in the forward direction. This result must therefore be a consequence of muscular anatomy and the tongue model's biomechanical properties.

### 3.3.1.3. *Movement amplitude during the consonantal closure*

It was noted in section 3.2, that, in all the V1-[k]-V2 simulations, the amplitude of the sliding movement of the tongue along the palate during the consonantal closure is mainly determined by the first vowel V1: the tongue slides over a distance of 5 mm for V1=[a], 3 mm for V1=[u], and around 2 mm for V1= [i].

In order to understand the origins of this phenomenon, different parameters were investigated: the amplitude and the orientation of the velocity vector just before consonant contact occurs, and the distances between nodes describing the tongue shape at the beginning of the consonantal closure and the virtual consonant tongue shape target (the shape it would assume without interference from the palate). Additional simulations of the [aka] sequence were also calculated with various transition times from [a] to [k], in order to change the velocity while keeping the target commands constant.

Considering all these simulations, no clear relation could be found between, on the one hand, the magnitude and direction of the velocity vector just before the contact and, on the other hand, the amplitude of the movement during the closure. The only systematic finding is related to distance between the tongue shapes at the beginning of the closure and at the consonant virtual target. This is illustrated by Figure 11, which shows these tongue shapes for [aka], [uku] and [iki] (from top to bottom). Considering the results depicted in Figures 6 through 8, it can be seen that the length of the sliding contact section of the movement is related to the distance between the position of the tongue when it first contacts the palate(C) and the position of the consonant's virtual target (V). In the case of vowels [u] and [a], starting from the vowel, the tongue moves first upward and forward until it hits the palate. From this time, the vertical movement becomes strongly constrained by the palatal contour. Since the tongue shape at the first point of contact is posterior to the virtual target shape of the consonant, the tongue continues to slide forward along the palate in the direction of the virtual consonant target, and the larger the distance between the two shapes, the larger the amplitude of the sliding movement.

**–INSERT FIGURE 11 AROUND HERE–**

For vowel [i], the first part of the movement is upward and backward. The movement in the vertical direction becomes strongly constrained when the tongue hits the palate, slightly in front of the consonantal target (recall that, in this case, the anterior target was used). Consequently, the tongue slides along the palate in the backward direction over a small distance.

The virtual target for the consonant is specified at the control level. The tongue shape at the beginning of the consonantal closure is the result of the tongue deformation from the vowel, which depends on muscular anatomy and biomechanical properties of the tongue (see section 3.3.1.2) and on the virtual target specified for V1 and [k]. We have shown that the amplitude of the movement during the consonantal closure depends on the distance between these two tongue shapes and on the interaction with the palate. Consequently, the amplitude of the movement during the closure is the result of a combination of effects related to the control (the virtual target sequence) and to biomechanical factors.

These observations can also explain the differences between our simulations and Mooshammer *et al.* (1995) measurements about the size of the loops in various V1-[k]-V2: while the general orientation of the loop is the same for each speaker, the amplitude of the sliding movement during the closure depends on speaker specific properties, at a control and at a physical level.

### 3.3.2. Reversal of loop direction through consonant target shifting

We have seen that in our simulations the direction of the loops is determined by the positions of the consonant and vowel target tongue shapes relative to each other. In this context, it should be interesting to determine the extent to which the generated patterns are sensitive to changes in the specified locations of the targets. More specifically, we are interested in conditions that would cause the directions of the loops to be reversed. Hence, additional simulations were generated, where the consonant target was moved in the direction opposite to the originally observed loop direction: for [aka] and [uku], the target was gradually moved backward to a position determined by increasing the recruitment of STY and decreasing the one of GGp; for [iki] it was moved forward by making the opposite changes in muscles recruitment.

**–INSERT FIGURE 12 AROUND HERE–**

Figure 12 shows, for [aka], the results of the simulation where the trajectory of the second node from the back becomes backward-oriented. The top panel shows in dotted line the virtual target used in [k] in the preceding simulations where forward oriented loops were observed, and in solid line the virtual target obtained by modifying the recruitment of STY and GGP to produce a tongue contour that is positioned at the place where the first reversal of a node trajectory could be observed (see bottom panel). The latter virtual target can be considered as a boundary within the vocal tract between two kinds of articulation for velar stops: for the virtual target tongue shapes that are more anterior than this boundary, the loops observed in [aka] will be forward oriented; for the virtual target tongue shapes that are more posterior, the loops will be backward oriented. Starting from the posterior target chosen for [k] in the preceding simulations, it took large changes in muscle commands to generate the differences in shape and to reverse the direction of the articulatory loop. As a consequence, the consonant target where reversion occurs is significantly different from the one used in the preceding simulations: the constriction is now essentially in the region of the soft palate, and not in the velo-palatal region, as usually observed for the velar stops /g/ and /k/ (Bothorel *et al.*, 1986). Similar results were found for [uku].

–INSERT FIGURE 13 AROUND HERE–

Figure 13 shows the result obtained for [iki], with a presentation identical to Figure 12. It can be seen that, contrary to [aka] a small forward shift of the consonant target, associated with very little changes in muscle commands, was enough to reverse the direction and the orientation of the loops, which are now forward directed and counter-clockwise oriented (as opposed to Figure 8, in which they are backward directed and clockwise oriented). The latter consonant target is still reasonable for a [k] articulated in a front vowel context (Bothorel *et al.*, 1986).

These results suggest that the forward direction of the looping patterns observed in the [a] and [u] contexts are very stable in the face of moderate changes in the consonant target location, while loop variability is likely to be observed in the [i] context, in which small perturbations of the target positions can reverse the loop direction and its orientation.

### 3.4. Effect of tongue-palate interaction.

It can be concluded from the preceding section that, according to our model, target locations, tongue muscle anatomy and biomechanics, together with the tongue palate interaction, may explain the existence, the direction, the orientation and the size of the loops. In this section, the effect of tongue-palate interaction will be discussed more specifically.

The influence of tongue-palate interaction on the articulatory trajectories can be illustrated quantitatively in our model by generating the same VCV sequences in a "virtual" vocal tract where the palate is removed. In this case, the consonant target can actually be reached, and the corresponding articulatory trajectories can be observed and compared to the simulation with the palate. The trajectories obtained for [aka] under the same conditions as above (Section 3.2), but with and without the palate, are shown in Figure 14. The top panel shows trajectories of four nodes on the dorsal contour of the tongue in the simulation of [aka] in a "virtual" vocal tract without palate. The palatal contour is shown as a reference with a solid line. The lower (solid) tongue contour represents the initial vowel configuration. The open symbols show the locations of the nodes at the following successive times: circles – when node 3 passes upward through the palatal contour (initial contact for the consonant when the palatal constraint is in effect), squares – when node 2 passes upward through the palatal contour, and triangles – just before node 3 passes downward through the palatal. The lower panel shows a superimposition of the trajectories simulated with (dashed line) and without (solid line) the palate. The same muscle commands were used for both simulations.

From the moment the tongue goes above the palate (circles in Figure 14, top half), the trajectories both of nodes 2 and 3 are oriented backward. The backward movement is more pronounced for node 3, due primarily to the conservation of volume constraint and the elastic properties of the model. In addition, the front part of the tongue (node 1), moves upward slightly after the central part of the tongue (nodes 2 and 3) has started to move downward (portions of the trajectories between the squares and triangles). In other words, the different nodes finish their upward movement at different times.

The lower part of Figure 14 shows that the four nodes initially first follow the same trajectory in both simulations. However, as would be expected, differences appear when the tongue first reaches the palatal contour. In the absence of palate, the tongue is free to continue its movement toward the virtual target without any limitation. From a little before the moment the upper part of the tongue goes beyond the palatal line (circles in the top half of the figure), its movement is no longer continuously upward and forward. Especially for the two middle nodes, an upward and backward movement occurs first; then the movement turns forward toward the virtual target location for the [k] (represented with the dotted line contour in the top panel). The backward movement is due to the fact that the force generated by the Styloglossus becomes larger than the force generated by the Posterior Genioglossus. According to our model of muscle force generation (Section 2.3.2.2), force variation is due to changes in macrofibers lengths induced by tongue deformation. This particular influence of the styloglossus could not be observed in simulations made with the palate, because, for the two middle nodes, the actions of the styloglossus and of the genioglossus, combined with the reaction force generated by palatal contact, resulted in a force in the forward direction. In the simulations without the palate, there is no reaction force; therefore, the net force acting on theses nodes as the upper part of the tongue goes beyond the palatal contour is first oriented in the rearward direction, before again becoming forward-oriented.

A comparison of the trajectories with the palate (dashed line, bottom half of Figure 14) with those without the palate (solid line) shows that, after initial contact, the trajectories without the palate are slightly more posterior than the trajectories with the palate for the 3 anterior nodes. Thus, in the simulations, the interactions between the tongue and the palate influence the trajectory shape. The adequacy of such predictions could be tested in the future with actual articulatory behavior and the use of a device that measures the pressure of the tongue against the palate.

Note however that, in both cases, the VC portion of the trajectory is located well behind the CV portion, and that the maximum distance in the mid-sagittal plane between theses two parts of the trajectory is not significantly modified by the presence or absence of palate. Therefore, in our model, the distance between the VC and the CV trajectories and the maximum size of the articulatory loops in the horizontal direction seem to depend only on tongue biomechanics and on virtual consonant target location, without any influence of the palate.

## 4. Conclusion

Simulations of VCV sequences (where C is a velar stop consonant) with a biomechanical model of the tongue have been presented. Both vowel and consonant gestures were controlled in terms of articulatory targets. Similar to observations on actual speakers, the VC and CV portions of the trajectories were somewhat curved and formed loops, even for symmetrical VCV sequences. The results seem to indicate that the presence and shape of the loops are strongly influenced by tongue biomechanics, including its muscular anatomy and contact with the palate. Contrary to suggestions by Löfqvist and Gracco (2002), biomechanics alone can be responsible for the trajectory curvature, and control of the entire trajectory based on a cost minimization principle does not seem to be necessary to explain these patterns. Of course, our results do not disprove Löfqvist and Gracco's (2002) hypothesis, since the control could act in combination with biomechanical factors. However, our simulations demonstrate that articulatory loops do not necessarily occur because entire articulatory trajectories are controlled in speech production. Our results support a more parsimonious theory of speech motor control, based on planning the target sequence and not the entire trajectory (Perrier et al., 1996a).

For [uku] and [aka], the results of the simulations that depict forward looping patterns are in good agreement with all the examples published in the literature. It was also shown for these two vowel contexts that in our simulations the direction and the orientation of the looping pattern is very resistant to changes in the position of the consonant target. Therefore, it seems that tongue biomechanics may explain the forward oriented loop trajectories that were observed for these sequences on a number of different speakers and in different languages, while the upper portion of the loop, is obviously influenced by interactions of the tongue with the palate.

Both for [iki] and [ika], in the first set of simulations, the model generated only backward movements. However, it was also shown for [iki] that a slight forward shift of the consonant target could induce a change in the loop direction and in its orientation. These results are consistent with the examples published in the literature. For example Houde (1967) observed for both sequences a small forward looping pattern, but Mooshammer *et al.*'s (1995) findings were slightly different. First, the latter authors did not observe any looping pattern for their two speakers during the production of [ika]. Second, for [iki], they confirmed Houde's observation, but they also noted that the velocity at the onset of the closure was oriented rearward for one of their two speakers and forward for the other speaker. Therefore, it seems reasonable to assume that there is a certain amount of variability between speakers, and perhaps also between languages, in the orientation and shape of articulatory trajectories, when the vowel preceding the velar consonant is [i]. This characteristic seems to be properly accounted for by our model. However, it should also be noted that in our simulations it was never possible to generate the kind of forward loop that Houde (1967) observed for [ika], which is "*superimposed, during contact, on the main rearward movement of the tongue*" (p. 129).

In conclusion, the simulations reported in the current paper suggest that, whatever the vowel context, the articulatory patterns observed in VCV sequences, where C is a velar stop consonant, are largely determined by tongue biomechanics. However, especially in the case of [iki] and [ika], where the orientation of these patterns seems to be quite unstable, it is probably necessary to take into account the potential role of other factors, such as the precise locations of the consonant and vowel targets and aerodynamics. Preliminary studies of the fluid-walls interaction in the vocal tract lead us to infer that aerodynamics could have an influence when V1=[i] (Perrier *et al.*, 2000).

In general, our findings partially support Hoole *et al.*'s (1998) suggestion that both aerodynamics and biomechanics probably contribute to the generation of the loops : "*Taken together, these observations suggest that the elliptical movement patterns found in speech must be put down to at least two factors: Firstly, aerodynamic factors operating in the vicinity of a consonantal constriction ; secondly, asymmetries in the muscle forces responsible for V-to-C and C-to-V movements.* " Hoole et al., 1998, p. 145). Our results may provide some answers to certain of Hoole's hypotheses. In particular, since in our simulations the commands patterns to all muscles are synchronized with each other, it may not be necessary to hypothesize temporal asymmetries in the muscle forces to account for the generation of the observed loops. In addition, compared to the biomechanics, aerodynamics may have a limited influence, especially in back vowel contexts.

## Acknowledgements

This work was supported by the CNRS, NSF and NIH (NIDCD Grant DC DC01925). The authors thank Torstein Are, Vincent Coisy, Frédéric Jolly and Jorge Vallejo, who contributed at different stages to the programming of the tongue model, Pierre Badin for the use of his X-ray data as a basis for the model and for valuable information about the 3D strain of the tongue, Christine Mooshammer and Susanne Fuchs for helpful comments at different stages of this work. Special thanks are also due to Anders Löfqvist and to the two reviewers Phil Hoole and Khalil Iskarous who provided extremely interesting and helpful comments.

**Footnotes**

[1] The Equilibrium Point Hypothesis (EPH) and its associated λ model are at the center of a number of controversies in the motor control literature (for example, see Feldman and Levin (1995), Gomi and Kawato (1996), Gottlieb (1998), Gribble et al., (1998)). Discussions about its possible use for the control of speech production have been presented in Laboissière *et al.* (1996), Perrier *et al.* (1996a, 1996b), and Sanguineti *et al.* (1998).

[2] Note that the amplitude of the sliding movement along the palate is identical for [ika] and [iki]. This is due simply to the fact that in both sequences the same targets were used for [i] and for [k]. Since a fairly long hold duration (100 ms) was specified for the consonant target command, no influence of the second vowel on the first part of the movement was to be expected. This is probably not very realistic, since, according to our hypothesis of speech production planning, the successive target commands should be selected specifically for each sequence, in order to optimize certain output constraints, such as economy of effort. This is a relatively minor detail that should not have a serious impact on the general conclusions drawn from our simulations.

**REFERENCES**

Adams, S.G., Weismer, G. and Kent, R.D. (1993). Speaking rate and speech movement velocity profiles. Journal of Speech and Hearing Research, 36, 41-54.

Badin, P., Gabioud, B., Beautemps, D., Lallouache, T.M., Bailly, G., Maeda, S., Zerling, J.P. and Brock, G. (1995). Cineradiography of VCV sequences: articulatory-acoustic data for a speech production model. *Proceedings of the 15th International Congress of Acoustics,* vol. IV (pp. 349-352), Trondheim, Norway.

Badin, P., Bailly, G., Revéret, L., Baciu, M., Segebarth, C. and Savariaux, C. (2002). Three-dimensional linear articulatory modeling of tongue, lips and face; based on MRI and video images. *Journal of Phonetics, 30*, 533-553

Bothorel, A., Simon, P., Wioland, F. and Zerling, J.-P. (1986). Cinéradiographie des voyelles et des consonnes du français. Institut de Phonétique, Université Marc Bloch, Strasbourg, France.

Bunton, K. and Weismer, G. (1994). Evaluation of a reiterant force-impulse task in the tongue. *Journal of Speech and Hearing Research, 37*, 1020-1031.

Cooper, S. (1953). Muscle spindles in the intrinsic muscles of the human tongue. *Journal of Physiology (London), 122*, 193-202.

Dang, J., and Honda, K. (1998). Speech synthesis of VCV sequences using a physiological articulatory model. P*roceedings of the 5th International Conference on Spoken Language Processing* (Vol. 5, pp. 1767-1770). Sydney, Australia.

Duck, F.A. (1990). *Physical properties of tissues: a comprehensive reference book.* London: Academic Press.

Fant G. (1960). *Acoustic Theory of Speech Production* . 216-228. The Hague: Mouton.

Feldman, A.G. (1966). Functional Tuning of The Nervous System with Control of Movement or Maintenance of a Steady Posture — II Controllable Parameters of the Muscles. *Biophysics*, *11*, 565-578.

Feldman, A.G. (1986). Once more on the Equilibrium-Point Hypothesis ( model) for motor control. *Journal of Motor Behavior*, *18 (1),* 17-54.

Feldman, A.G., and Orlovsky, G.N. (1972). The influence of different descending systems on the tonic stretch reflex in the cat. Experimental Neurology, 37, 481-494.

Feldman, A.G. and Levin, M.F. (1995). Positional frames of reference in motor control: Origin and use. *Behavioral and Brain Sciences*, *18 (4)*, 723-806.

Flanagan, J.R., Ostry, D.J. and Feldman, A.G. (1990). Control of human jaw and multi-joint arm movements. In G.E. Hammond (Ed.), *Cerebral control of speech and limb movements* (pp.29-58), Amsterdam, The Netherlands: Elsevier Science Publishers B.V. (North-Holland).

Fowler, C.A. (1980). Coarticulation and theories of extrinsic timing. *Journal of Phonetics*, *8*, 113-133.

Fuchs, S., Perrier, P. and Mooshammer, C. (2001). The role of the palate in tongue kinematics: an experimental assessment in VC sequences from EPG and EMMA data. *Proceedings of Eurospeech 2001* (pp. 1487-1490). Aalborg, Denmark.

Fung, Y.C. (1993) *Biomechanics : mechanical properties of living tissues.* New York: Springer-Verlag

Gomi, H., and Kawato, M. (1996). Equilibrium-point control hypothesis examined by measured arm-stiffness during multi-joint movement. *Science, 272*, 117-120.

Gottlieb, G.L. (1998). Rejecting the Equilibrium Point Hypothesis, *Motor Control*, *2*, 10-12

Gribble, P.L., Ostry, D.J., Sanguineti, V. and Laboissiere, R. (1998). Are complex control signals required for human arm movement? *Journal of Neurophysiology, 79(3)*, 1409-1424

Hashimoto, K. and Suga, S. (1986). Estimation of the muscular tensions of the human tongue by using a three-dimensional model of the tongue. *Journal of Acoustic Society of Japan (E)*, *7 (1)*,39-46.

Hoole, P., Munhall, K. and Mooshammer, C. (1998). Do air-stream mechanisms influence tongue movement paths? *Phonetica*, *55*, 131-146

Honda, K. (1996). The organization of tongue articulation for vowels. *Journal of Phonetics, 24(1)*, 39-52

Houde, R.A. (1967). *A study of tongue body motion during selected speech sounds*, Unpublished Ph.D. Dissertation., University of Michigan.

Husson, R. (1950). *Etude des phénomènes physiologiques et acoustiques fondamentaux de la voix chantée.* Unpublished Ph.D. Dissertation, University of Paris

Kakita, Y., Fujimura, O., and Honda, K. (1985). Computation of mapping from muscular contraction patterns to formant patterns in vowel space. In V.A. Fromkin (Ed.), *Phonetic Linguistics* (pp. 133-144). Orlando, Florida: Academic Press.

Keating, P. (1993). Fronted velars, palatalized velars, and palatals. *Phonetica, 50*, 73-101.

Kent, R. and Moll, K. (1972). Cinefluorographic analyses of selected lingual consonants. *Journal of Speech and Hearing Research, 15*, 453-473.

Kiritani, S., Miyawaki, K. and Fujimura, O. (1976). A computational model of the tongue. *Annual Report of the Research Institute of Logopedics and Phoniatrics, 10*, 243-252, Tokyo University.

Laboissière, R., Ostry, D.J. and Feldman, A.G. (1996). The control of multi-muscle systems: human jaw and hyoid movements. *Biological Cybernetics, 74(3)*, 373-384.

Löfqvist, A. and Gracco, V.L. (1994). Tongue body kinematics in velar stop production: influences of consonant voicing and vowel context, *Phonetica, 51*, 52-67.

Löfqvist, A. and Gracco, V.L. (1997). Lip and jaw kinematics in bilabial stop consonant production, *Journal of Speech Language and Hearing Research*, *40*, 877-893.

Löfqvist, A. and Gracco, V.L. (2002). Control of oral closure in lingual stop consonant production, *Journal of the Acoustical Society of America*, *111(6)*, 2811-2827.

Marhefka, D.W. and Orin, D.E. (1996). Simulations of contact using a non-linear damping model. *Proc. of IEEE International Conference on Robotics and Automation*, *Vol. 2*.(pp. 1662-1668) (Minneapolis, MN).

McClean, M.D. and Clay, J.L. (1995). Activation of lip motor units with variations in speech rate and phonetic structure. *Journal of Speech and Hearing Research, 38*, 772-782.

Min, Y., Titze, I. and Alipour, F. (1994). Stress-Strain Response of the Human Vocal Ligament. *NCVS Status and Progress Report, 7*, 131-137.

Mooshammer, C., Hoole, P., and Kühnert, B. (1995). On loops. *Journal of Phonetics*, *23*, 3-21.

Nelson, W.L. (1983). Physical principles for economies of skilled movements. *Biological Cybernetics, 46*, 135-147.

Netter, F.H. (1989). *Atlas of human anatomy*. CIBA-GEIGY Corporation editor.

Ohala, J. (1983). The Origin of Sound Patterns in Vocal-Tract Constraints. In P.F. MacNeilage (ed.), *The Production of Speech* (pp. 189-216). New York: Springer-Verlag.

Ohayon, J., Usson, Y., Jouk, P.S. and Cai, H. (1999). Fibre orientation in human fetal heart and ventricular mechanics: A small perturbation analysis. *Computer Methods in Biomedical Engineering*, *2*, 83-105.

Öhman, S.E.G. (1966). Coarticulation in VCV utterances: spectrographic measurements, *Journal of the Acoustical Society of America*, *39*, 151-168.

Ostry, D.J. and Munhall, K.G. (1985). Control of rate and duration of speech movements. *Journal of the Acoustical Society of America*, *77*, 640-648.

Matthies, M.L., Perrier, P., Perkell, J.S. and Zandipour, M. (2001). Variation in speech movement kinematics and temporal patterns of coarticulation with changes in clarity and rate. *Journal of Speech Language and Hearing Research, 44 (2)*, 340-353.

Payan, Y. and Perrier, P. (1997). Synthesis of V-V Sequences with a 2D Biomechanical Tongue Model Controlled by the Equilibrium Point Hypothesis. *Speech Communication, 22, (2/3)*, 185-205

Perkell, J.S. (1969). *Physiology of speech production: results and implication of a quantitative cineradiographic study*. Cambridge, Massachusetts: MIT Press.

Perkell, J.S. (1974). *A physiologically oriented model of tongue activity in speech production*. Unpublished Ph.D. Dissertation. Cambridge, MA: Massachusetts Institute of Technology.

Perkell, J.S. and Matthies, M.L. (1992). Temporal measures of anticipatory labial coarticulation for the vowel [u]: within- and cross-subject variability. *Journal of the Acoustical Society of America, 91*, 2911-2925.

Perkell, J.S., Svirsky, M.A., Matthies, M.L. and Manzella, J. (1993). On the use of Electro-Magnetic Midsagittal Articulometer (EMMA) systems. *Forschungsberichte des Instituts für Phonetik und Sprachliche Kommunikation der Universität München, 31*, 29-42. University of Munich, Germany.

Perkell, J.S. (1996). Properties of the tongue help to define vowels categories : Hypotheses based on physiologically oriented modeling. *Journal of Phonetics, 24*, 3-22

Perkell, J.S., Guenther, F.H., Lane, H., Matthies, M.L., Perrier, P., Vick, J., Wilhelms-Tricarico, R., and Zandipour, M. (2000). A theory of speech motor control and supporting data from speakers with normal hearing and with profound hearing loss. *Journal of Phonetics, 28(3)*, 233-272.

Perkins, W.H. and Kent, R.D. (1986). *Functional anatomy of speech, language and hearing*. Needham Heights, MA : Allyn and Bacon.

Perrier, P., Loevenbruck, H. and Payan, Y. (1996)a. Control of tongue movements in speech: the Equilibrium Point Hypothesis perspective. *Journal of Phonetics*, *24*, 53-75,

Perrier, P., Ostry, D.J. and Laboissière, R. (1996b). The Equilibrium Point Hypothesis and its application to speech motor control. *Journal of Speech and Hearing Research*, *39*, 365-378.

Perrier, P., Payan, Y., Perkell, J., Zandipour, M., Pelorson, X., Coisy, V. and Matthies, M. (2000). An attempt to simulate fluid-walls interactions during velar stops. In *Proceedings the 5th Speech Production Seminar and CREST Workshop on Models of Speech Production* (pp. 149_152). Kloster Seeon, Bavaria.

Porter, R. (1966) Lingual mechanoreceptors activated by muscle twitch. *Journal of Physiology (London), 183*,101-111

Sanguineti, V., Laboissière, R. and Payan, Y. (1997). A control model of human tongue movements in speech. *Biological Cybernetics*, *77(11)*, 11-22

Sanguineti, V., Laboissière, R. and Ostry, D.J. (1998). A dynamic biomechanical model for the neural control of speech production. *Journal of the Acoustical Society of America, 103*, 1615-1627

Stone, M., Goldstein, M., H., and Zhang Y. (1997). Principal component analysis of cross sections of tongue shapes in vowel production. *Speech Communication, 22*, 173-184

Straka, G. (1965). *Album phonétique*. Québec, Canada : Les Presses de l'Université Laval.

Takemoto, H. (2001). Morphological analyses of the human tongue musculature for three-dimensional modeling. *Journal of Speech, Language, and Hearing Research*, *44*, 95-107.

Trulsson, M and Essick, G.K. (1997). Low-threshold mechanoreceptive afferents in the human lingual nerve. *J. Neurophysiology*, *77*, 737-748.

Van den Berg, J. (1958), Myoelastic-aerodynamic theory of voice production. *Journal of Speech and Hearing Research, l*, 227-244

Walker, L.B. and Rajagopal, M.D.(1959). Neuromuscular spindles in the human tongue. *Anatomical Record*, *133*, 438.

Wilhelms-Tricarico, R. (1995). Physiological modeling of speech production: Methods for modeling soft-tissues articulators. *Journal of the Acoustical Society of America, 97(5)*, 3085-3098.

Wood, S. A. J. (1979). A radiographic analysis of constriction locations for vowels. *Journal of Phonetics* 7, 25-43.

Zienkiewicz, O.C. and Taylor, R.L. (1989). *The Finite Element Method. Basic Formulation and Linear Problems.* Maidenhead, UK: MacGraw-Hill Book Company Limited,

**Table I:** Timing of the commands for the VCV sequences

|  | **Duration (ms)** |
|:---:|:---:|
| Vowel Hold time | 150 |
| Vowel-to-[k] Transition Time | 30 |
| [k] Hold Time | 100 |
| [k]-to-Vowel Transition Time | 30 |
| Vowel Hold Time | 150 |

# Figure Captions

**Figure 1:** Mesh of the 2D Finite Element tongue model in its rest position. The external vocal tract contours were extracted from X-Ray data collected on the reference speaker PB.

**Figure 2**: Representation of the seven muscles taken into account in the model. The bold lines represent macro-fibers, over which the global muscle force is distributed. The gray shaded quadrilaterals are selected elements within the FE structure, whose mechanical stiffness increases with muscle activation.

**Figure 3:** Tongue deformations associated with muscle activations. The dotted line represents the contour of the tongue at rest. Units of X and Y axes are in mm.

**Figure 4:** Tongue shapes for the vowel targets used in the simulations. The dotted line represents the contour of the tongue at rest. Units of X and Y axes are in mm.

**Figure 5:** Tongue shapes for the virtual consonant targets used in the simulations. Top panel: posterior target; middle panel: anterior target; in these two panels, the dotted line represents the contour of the tongue at rest; bottom panel: enlarged view of the tongue contours in the palatal region; dotted line: posterior target ; solid bold line: anterior target. Units of X and Y axes are in mm.

**Figure 6:** Trajectories of four nodes on the dorsal contour of the tongue in the simulation of [aka]; top panel: general sagittal view; bottom panel: close up in the palatal region. The solid tongue contour represents the initial vowel configuration; the dotted line contour corresponds to the consonant configuration just before release; for each trajectory, the starting point is marked with a small open circle on the solid line tongue contour. Units of X and Y axes are in mm.

**Figure 7:** Trajectories of four nodes on the dorsal contour of the tongue in the simulation of [uku]; top panel: general sagittal view; low panel: close up in the palatal region. The solid tongue contour represents the initial vowel configuration; the dotted line contour corresponds to the consonant configuration just before release; for each trajectory, the starting point is marked with a small open circle on the solid line tongue contour. Units of X and Y axes are in mm.

**Figure 8**: Trajectories of four nodes on the dorsal contour of the tongue in the simulation of [iki]; top panel: general sagittal view; low panel: close up in the palatal region. The solid tongue contour represents the initial vowel configuration; the dotted line contour corresponds to the consonant configuration just before release; for each trajectory, the starting point is marked with a small open circle on the solid line tongue contour. Units of X and Y axes are in mm.

**Figure 9**: Trajectories of four nodes on the dorsal contour of the tongue in the simulation of [ika]; top panel: general sagittal view; low panel: close up in the palatal region. The solid tongue contour represents the initial vowel configuration; the dotted line contour corresponds to the consonant configuration just before release; for each trajectory, the starting point is marked with a small open circle on the solid line tongue contour. Units of X and Y axes are in mm

**Figure 10:** Trajectories of a node on the dorsal tongue contour of the tongue (the second node from the back on figures 6 to 9) in the asymmetrical sequences; top panel: [ik]-V2 sequences; middle panel: [ak]-V2 sequences; bottom panel: [ak]-V2 sequences, where V2 is one of vowels [i] (dashed-dotted lines), [a] (solid lines) and [u] (dotted lines). The solid arrows show the directions of movements in the closing phase toward the consonant. Units of X and Y axes are in mm.

**Figure 11**: Close up in the palatal region of the tongue shapes (bold line, labeled C) at the beginning of the consonantal closure for [aka], [uku] and [iki]. The dotted line represents the contour of the tongue at the virtual consonant target (labeled V). The solid line represents the palatal contour. Units of X and Y axes are in mm.

**Figure 12**: Generation of backward oriented loops for [aka] through target shifting. Top panel: original (dotted line) and modified (bold line) virtual consonant targets; lower panel: Trajectories of four nodes on the dorsal contour of the tongue in the simulation of [aka] with the modified consonant target: close up in the palatal region. The solid tongue contour represents the initial vowel configuration; the dotted line contour corresponds to the consonant configuration just before release; for each trajectory, the starting point is marked with a small open circle on the solid line tongue contour. Units of X and Y axes are in mm.

**Figure 13**: Generation of forward oriented loops for [iki] through target shifting. Top panel: original (dotted line) and modified (bold line) virtual consonant targets; lower panel: Trajectories of four nodes on the dorsal contour of the tongue in the simulation of [iki] with the modified consonant target: close up in the palatal region. The solid tongue contour represents the initial vowel configuration; the dotted line contour corresponds to the consonant configuration just before release; for each trajectory, the starting point is marked with a small open circle on the solid line tongue contour. Units of X and Y axes are in mm.

**Figure 14**: Top panel: Trajectories of four nodes on the dorsal contour of the tongue in the simulation of [aka] in a "virtual" vocal tract without palate. The palatal contour is represented as a reference. The solid tongue contour represents the initial vowel configuration. The open symbols show the locations of the nodes at the following successive times: circles – when node 3 passes upward through the palatal contour (initial contact when the palatal constraint is in effect), squares – when node 2 passes upward through the palatal contour, and triangles –

just before node 3 passes downward through the palatal contour (consonant release when the palatal constraint is in effect). The dotted contour corresponds to the virtual target of the consonant. The starting point is marked with a small filled circle on the solid line tongue contour. Lower panel: superimposition of the trajectories simulated with (dashed line) and without (solid line) palate. Units of X and Y axes are in mm

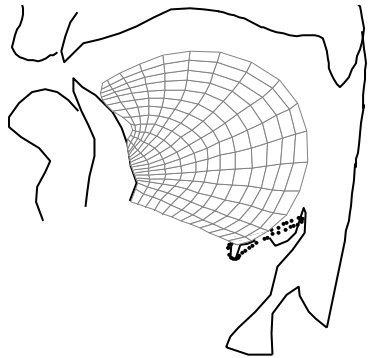

**Figure 1**

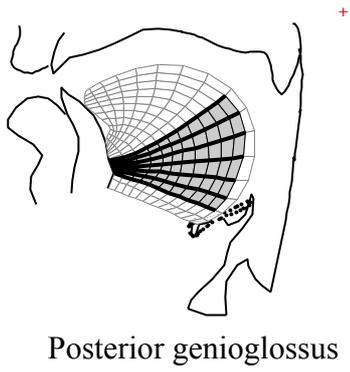
Posterior genioglossus

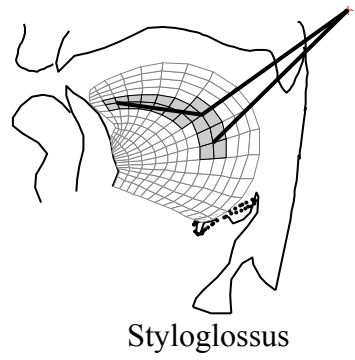
Styloglossus

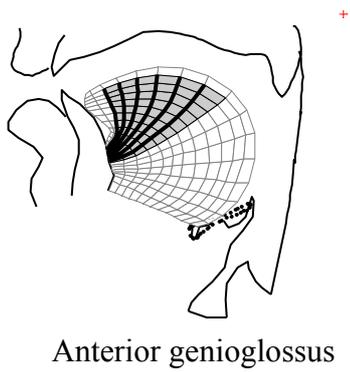
Anterior genioglossus

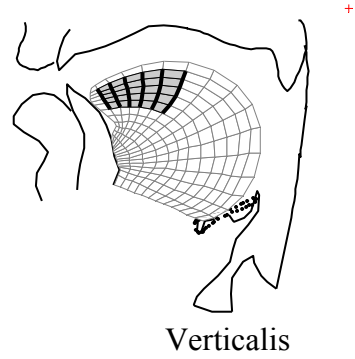
Verticalis

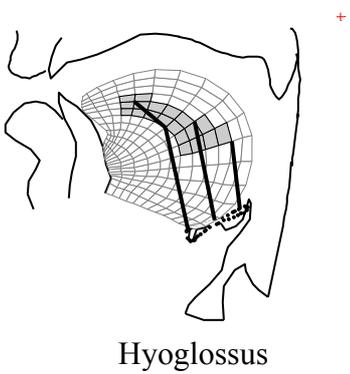
Hyoglossus

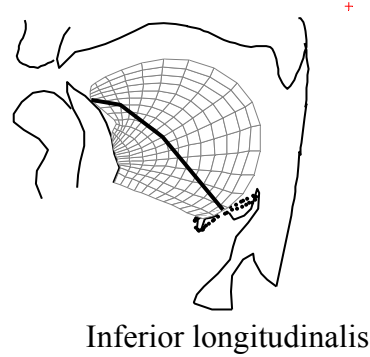
Inferior longitudinalis

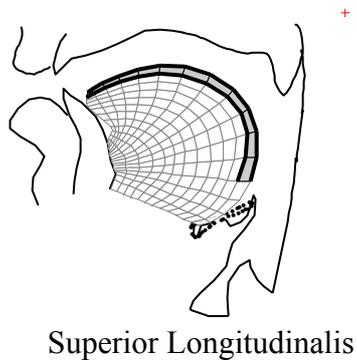
Superior Longitudinalis

**Figure 2**

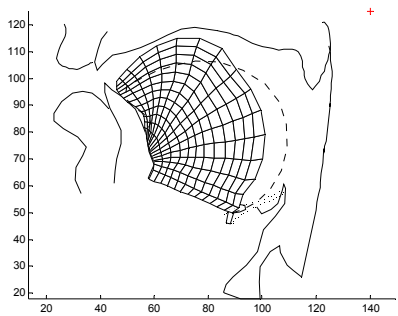
Posterior genioglossus

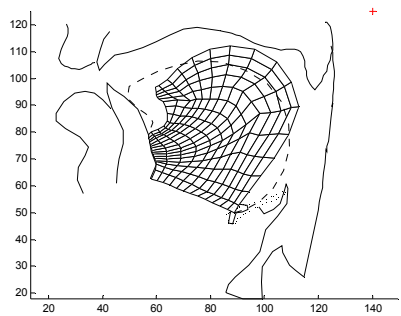
Styloglossus

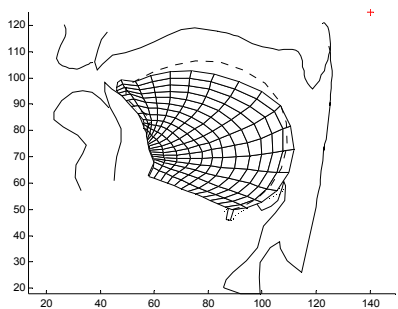
Anterior genioglossus

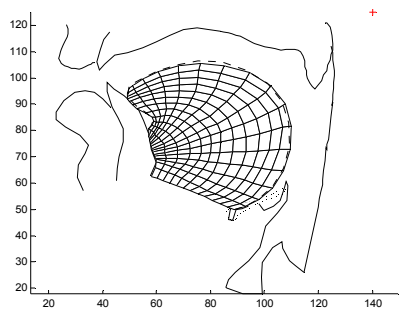
Verticalis

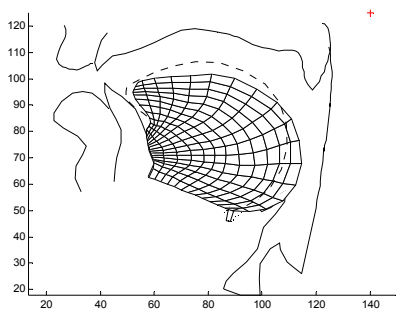
Hyoglossus

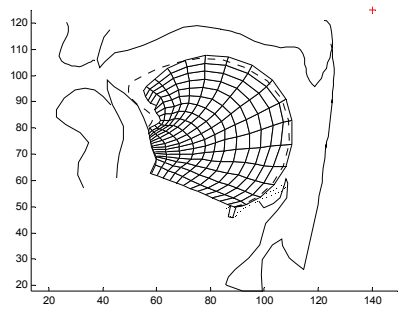
Inferior longitudinalis

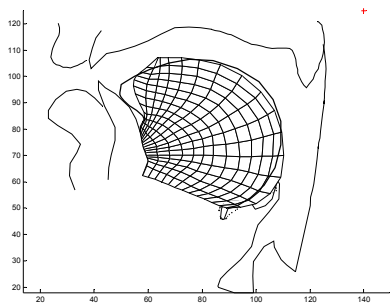
Superior Longitudinalis

**Figure 3**

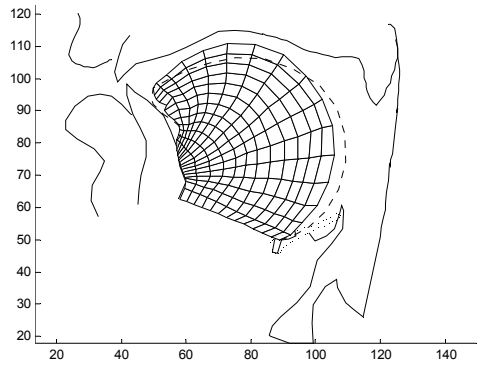

Vowel [i]

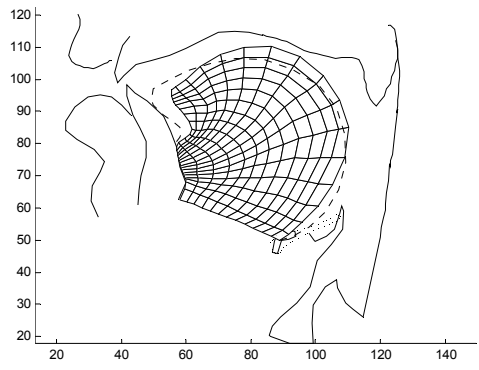

Vowel [u]

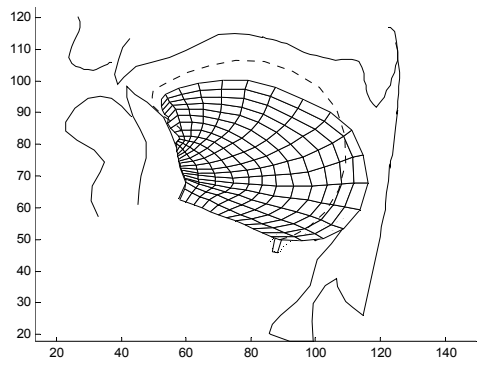

Vowel [a]

**Figure 4**

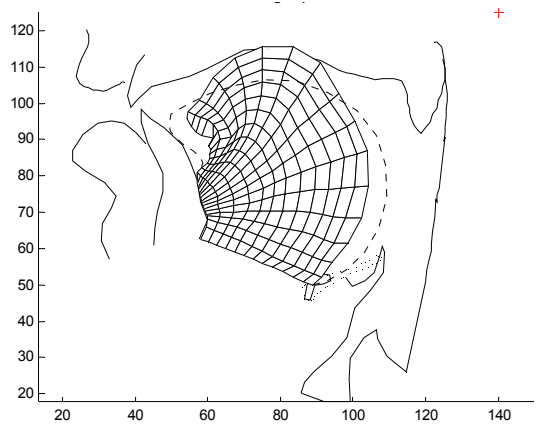

Anterior target for [k]

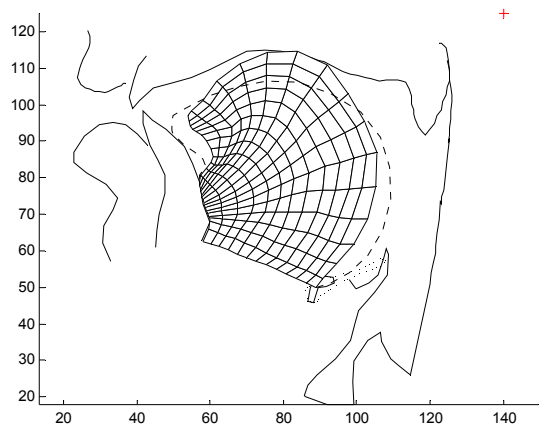

Posterior target for [k]

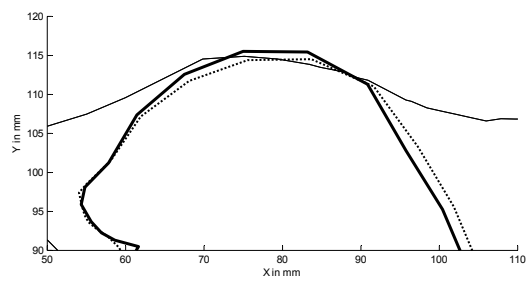

Enlarged view in the palatal region

**Figure 5**

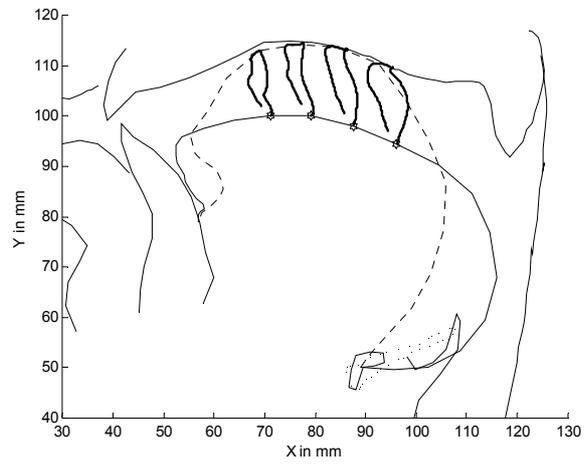

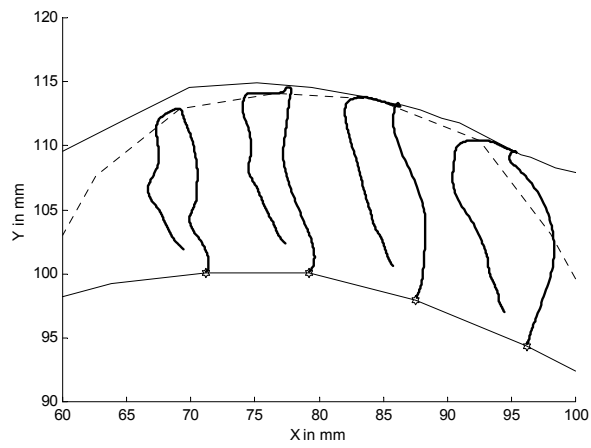

**Figure 6**

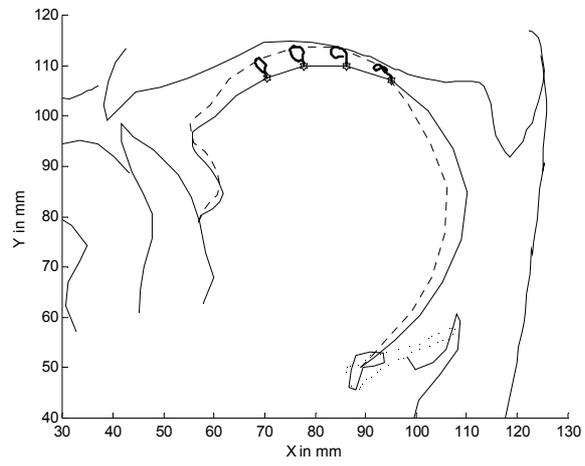
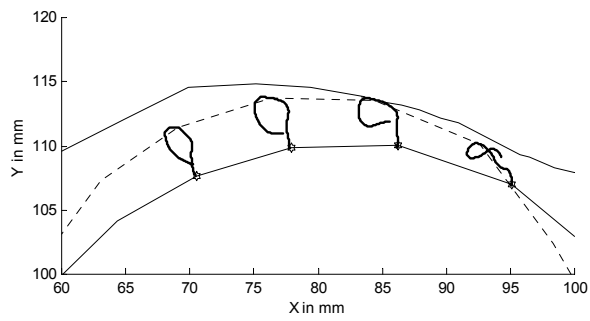

**Figure 7**

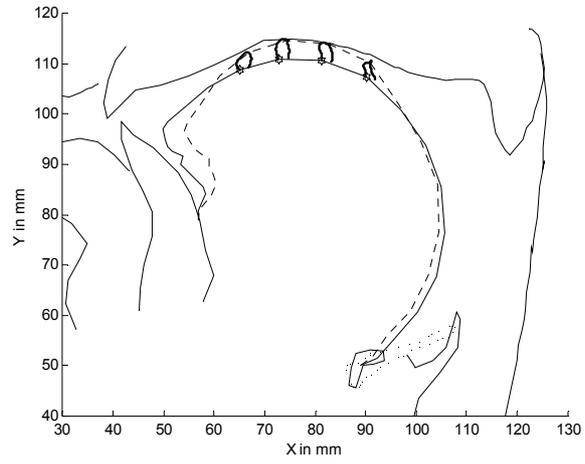

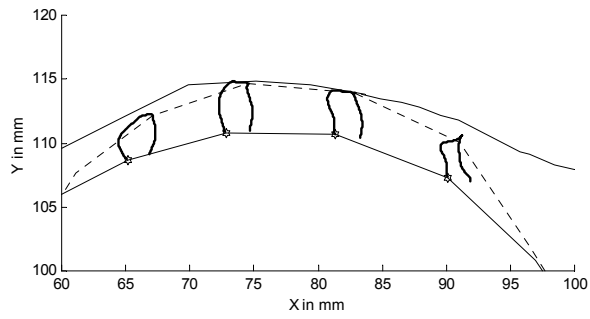

**Figure 8**

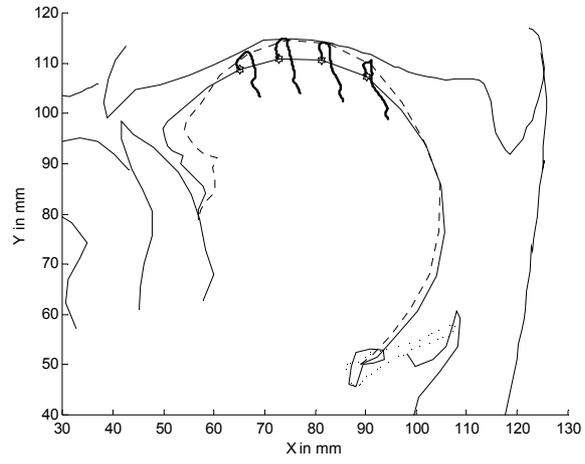
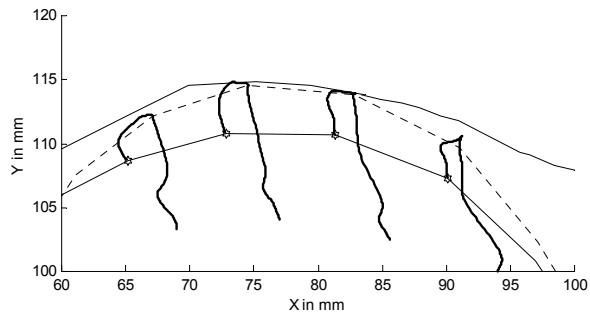

**Figure 9**

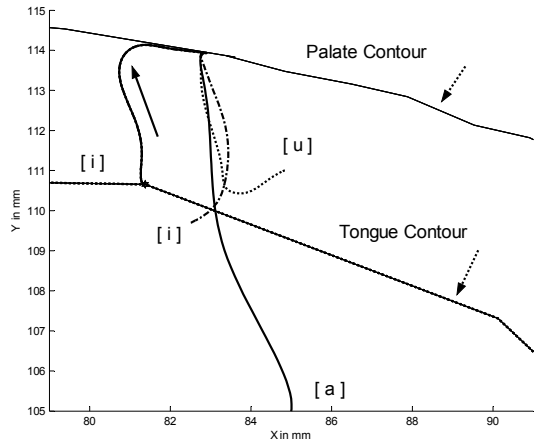

[ik]-V2

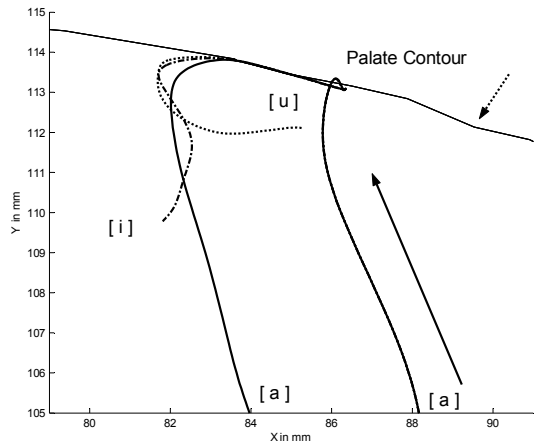

[ak]-V2

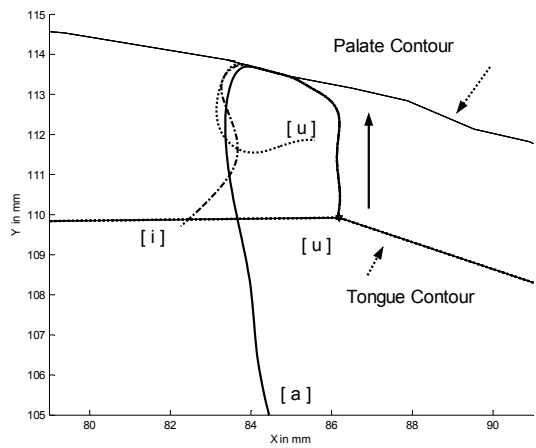

[uk]-V2

**Figure 10**

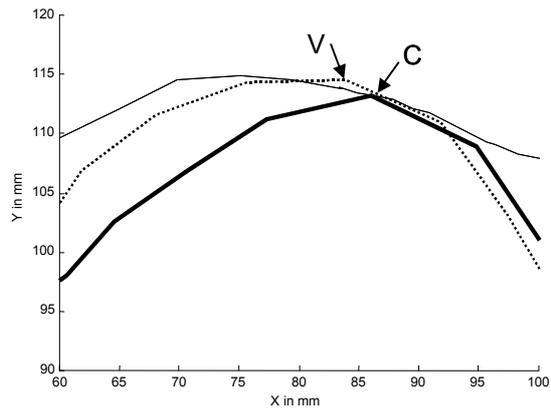

in [aka]

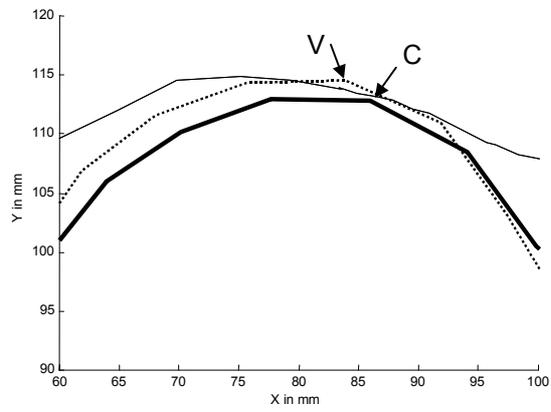

in [uku]

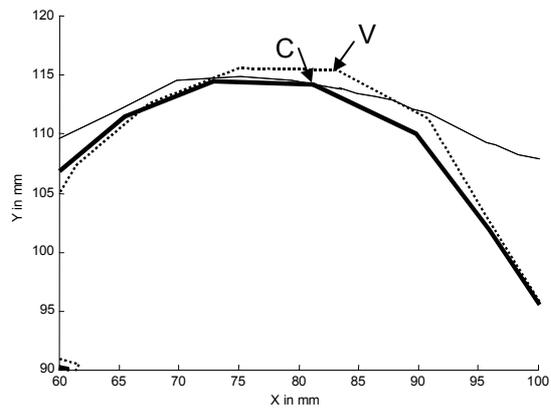

in [iki]

**Figure 11**

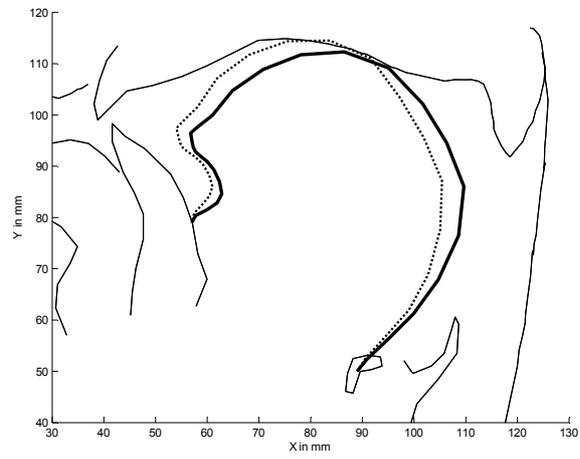

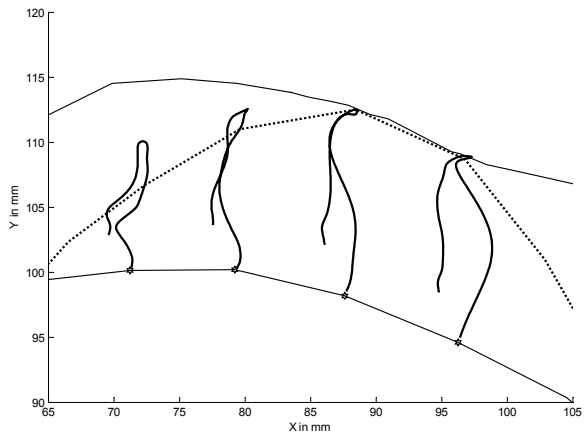

**Figure 12**

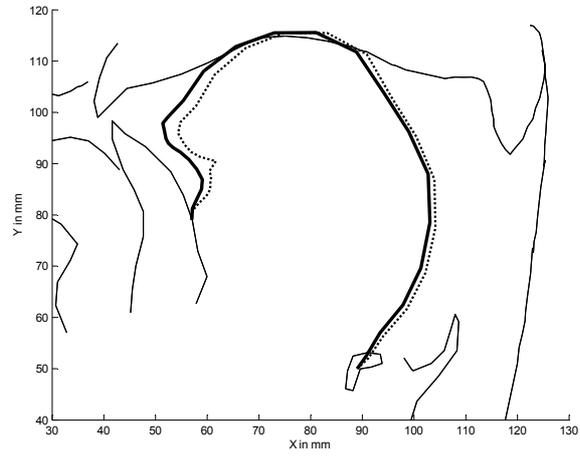

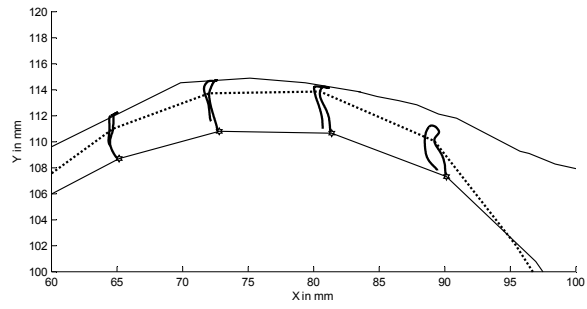

**Figure 13**

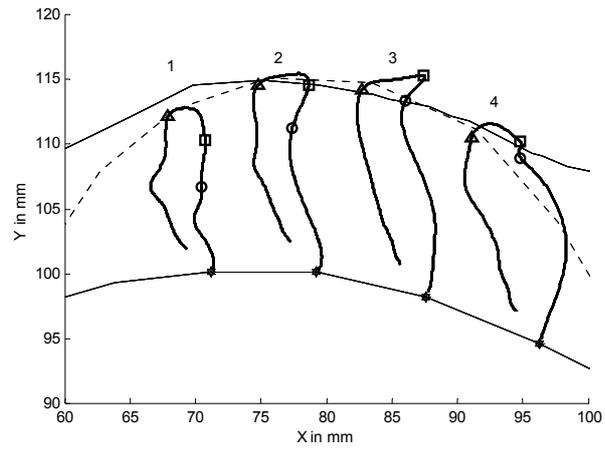

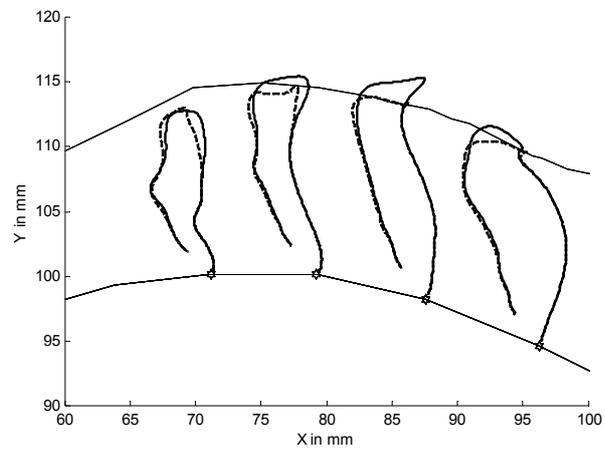

**Figure 14**